\documentclass{aa}
\usepackage{epsfig}

\def \etal {et al.\ } 
\def \xray {\hbox{X--ray }} 
\def \rc {r_{\rm c}} 
\def \Lcl {\Lambda_{\rm cl}} 
 
\def \rdc {r_{\rm 200}}
\def \rosat {\hbox{\it ROSAT }} 
\def \xmm {\hbox{\it XMM-Newton }} 
\def \cha {\hbox{\it Chandra }}
\def \clone {\hbox{CL~J0533-2411 }} 
\def \cltwo {\hbox{EIS~J0533-2412 }} 
\def \xmmone {\hbox{XMMU~J053344.4-241048 }} 
\def \xmmtwo {\hbox{XMMU~J053337.8-241250 }} 
\def \betamodel {\hbox{$\beta$--model }}

\begin{document}
\title{The origin of X-ray emission of two distant ($z>1$) cluster candidates
with XMM-Newton}

\author{ D.M. \,Neumann \inst{1} 
\and M. \,Arnaud\inst{1}
\and C. \,Benoist\inst{2} 
\and L. \,da Costa\inst{3} \and H.E. 
\,J{\o}rgensen\inst{4} \and L.F. \,Olsen\inst{4} \and S. \,Bardelli\inst{5} \and
E. \,Zucca\inst{5} \and S. \,Arnouts\inst{6} \and A. \,Biviano\inst{7}
\and M. \,Ramella\inst{7}
 } 
\offprints{D.M. \,Neumann, \email{ddon@cea.fr}}
\institute{ CEA/Saclay DSM/DAPNIA/SAp, L'Orme des Merisiers, 91191 Gif-sur-Yvette, France
\and
Observatoire de la C\^ote d'Azur, CERGA, BP229, Nice, cedex 4, France
\and
ESO, Karl-Schwarzschild-Straße 2, 85748 Garching, Germany
\and
Astronomical Observatory, Juliane Maries Vej 30,
DK-2100 Copenhagen, Denmark
\and 
INAF - Osservatorio Astronomico di Bologna
     via Ranzani 1
     40127 Bologna, Italy
\and
Laboratoire d'Astrophysique de Marseille, Traverse  du Siphon - BP 8, 
13376 Marseille Cedex 12, France
\and
INAF - Osservatorio Astronomico di Trieste
                         via G.B. Tiepolo 11, 34131, Trieste, Italy
}

\abstract{We present here a study of \xmm data of two distant
galaxy cluster candidates.  One of these was discovered
serendipitously in near infrared data, \object{CL~J0533-2411}, the other one
corresponds to the cluster \object{EIS~J0533-2412} part of the EIS cluster 
survey.  The estimated redshift of \clone is
$z=1.2-1.7$.  \cltwo is a rich system ($\Lcl=299$), with a
spectroscopically confirmed redshift of $z=1.3$.  Both galaxy
concentrations show firm X-ray detections, located within $30^{\prime\prime}$ of
their optical center.  However, we cannot resolve the sources with
\xmm.  If the X-ray emission originates from the X-ray emitting
intra-cluster medium (ICM) it would be extremely concentrated which is rather
unlikely (core radii below $14~{\rm h_{65}^{-1}~kpc}$ and $40~{\rm
h_{65}^{-1}~kpc}$, respectively).  We argue that the X--ray sources
are more likely AGN members of the galaxy concentrations.  We set an
upper limit for the bolometric luminosity of a hot ICM in the range
$\sim 0.7-2.1 \times 10^{44}~{\rm h_{65}^{-2}~erg/s}$ for \clone,
depending on the exact redshift.  For \cltwo the limit is $L_{\rm
bol}= (6.2\pm1.4) \times 10^{43}~{\rm h_{65}^{-2}~erg/s}$.  We
interpret our result in the following way: \cltwo (and possibly
\clone) are proto-clusters and show matter overdensities before
collapse, which explains the low significance of extended X-ray
emission.  }

\authorrunning{D.M. Neumann \etal} \titlerunning{Distant cluster
candidates with XMM-Newton}
\maketitle

\section{Introduction}

Cluster evolution with time provides essential information on the
physics of structure formation and can set strong constraints on the
cosmological parameters.  In spite of remarkable progresses in that
field in the recent years (e.g. Rosati, Borgani \& Norman \cite{rbn}
and references therein), the redshift range $z>1$ remains
largely unexplored.  

The results presented in this paper are part of an ongoing
comprehensive effort to identify and study clusters at different
epochs using as starting point the ESO Imaging Survey (EIS -- Nonino
\etal~\cite{nonino}; Prandoni \etal~\cite{prandoni}; Benoist 
\etal~\cite{benoist99}) and the subsequent cluster candidate compilation,
which was built using a matched filter technique in the I band (Olsen
\etal~\cite{ols99}; Scodeggio \etal~\cite{sco99}).  These original
data were complemented by multi-band optical/infrared imaging data to
derive photometric redshifts.  Combining these estimates with the
positional information allows to search for concentrations in the 3D
space and to select the most likely clusters (da Costa
\etal~\cite{dac99}).  First VLT spectroscopic observations confirmed the
reality of three candidates in the redshift range $z=0.8-1.3$
(Benoist \etal \cite{ben02}).

We present here \xmm follow-up of two cluster candidates at $z>1$. 
The first object, \object{CL~J0533-2411}, was detected serendipitously
in the J/K follow up of \cltwo with SOFI/NTT, as an overdensity in the
projected distribution of galaxies with similar J-K colors.  Based on
the I, J and K-band data the estimated redshift of this cluster
candidate is $z=1.2-1.7$.  The other object is the system
\object{EIS~J0533-2412}, which is supposed to be a rich system based
on the matched filter parameter $\Lambda_{\rm cl}$ ($\Lambda_{\rm
cl}=299$).  VLT/FORS2 spectroscopic observations strongly suggest 
that this
galaxy concentration is a true physical structure at $z=1.3$
(Benoist \etal~\cite{ben02}). The group in this field at z=0.8 as 
indicated by
Benoist \etal~\cite{ben02} is scattered through the field and does not 
coincide with the matched filter position of the \cltwo cluster. Detailed
spectral information of the cluster galaxies will be presented in a forthcoming
paper by J{\o}rgensen et al. \cite{jorg}.

X--ray observations are a key to understand the nature of these galaxy
systems, which at such high redshift are not necessarily fully
collapsed and relaxed objects.  They could be filaments,
proto-clusters or more or less virialised clusters.  The detection of
extended X--ray emission from hot gas, heated by gravitational
collapse, would trace the virialized portion of these distant  and young
systems.  

The paper is structured in the following way.  The data are described
in Sect.~ 2.  In Sect.~3 we present the data reduction and the results
of our analysis.  We examine the significance and extent of the \xray
emission at the location of both cluster candidates, the physical
origin of the emission and put constraints on the ICM \xray
luminosities.  We discuss the physical nature of the cluster
candidates in Sect.~4 and conclude in Sect.~5.

Throughout this paper we use $H_0 = 65~h_{65}$~km/s/Mpc, and $\Omega_m =
0.3$, $\Omega_\Lambda = 0.7$.  In this cosmology, one arcminute corresponds
to $0.541$~Mpc and $0.547$~Mpc at a redshift of $z = 1.3$ and $z=1.7$, 
respectively.

\section{The \xmm data}

Both clusters lie sufficiently close together (2 arcmin distance) to
be observed in one \xmm observation.  The data were taken in two
exposures: $64.9$~ksec during revolution $332$ and $34.5$~ksec during
revolution $335$.  The total exposure time was longer than originally
scheduled since the pn-camera was not operational at that time.  Thus
we present here only data based on EPIC-MOS cameras.  

\section{Data analysis}

For our analysis we use the official \xmm software (SAS version
5.2).  The data presented below are corrected for vignetting
effects using the photon weighting method (Arnaud
\etal~\cite{arnaud01}; Majerowicz \etal~\cite{majero}).

\subsection{Background Screening}

{\it XMM-Newton} observations suffer from time periods of high background
due to soft protons from solar flares.  During those flares the
background level can increase by several orders of magnitude as
compared to the quiescent background level and significantly degrades
the signal to noise ratio.

We bin the MOS-data in time intervals of 100~sec in the high energy
band 10-12~keV band, where the emission is dominated by the particle
induced background.  We screen the data for flare periods by applying
a rejection threshold in this energy  band of $>15$ counts.  After such
cleaning, the effective exposure time is $83$~ksec and $84$~ksec for the
MOS1 and MOS2 cameras, respectively.

\subsection{Source detection and morphology}

\begin{figure*}[ht]
\begin{center}
\vspace*{16cm}
\caption{The contours of 
the X-ray emission in sigmas of significance.  Lowest contour and
stepwidth 1 sigma.  The contours are overlaid over the SOFI/NTT image
taken in the K--band. The squares indicate galaxies with  $J-K=2-2.4$
corresponding to early type galaxies with redshifts approximately beyond 1.0.
The three spectroscopically confirmed members of \cltwo are indicated by 
circles. The crosses show the center
of the galaxy concentrations.}
\label{fig:K}
\end{center}
\end{figure*}

In order to enhance the signal-to-noise ratio we only take into
account data in the energy band 0.3-3.0~keV. We extract the MOS1\&2
image with a pixel size of $1.1\arcsec$ times $1.1\arcsec$, which
corresponds to the physical size of the detector pixels.  The X-ray
signal at the location of both \clone and \cltwo appear very faint and
we therefore apply first a detection algorithm, which gives the
significance of the signals.  We use the code described in Neumann \&
B\"ohringer (\cite{neu97}), which assumes Poisson statistics and uses
a Gauss filter.  We apply this algorithm with a Gauss filter size of
$\sigma=5.5\arcsec$ and an overall background estimate of 0.004
cts/s/arcmin$^2$ in the 0.3--3.0~keV energy range.

\begin{table}[b]
\caption[]{ Object position}
\begin{flushleft} 
\begin{tabular}{lll}
\hline 
\hline 
Objects & RA (J2000.0) & DEC (J2000.0) \\
\hline 
\clone & $05^{h}33^{m}44.0^{s}$& $-24\degr11\arcmin01\arcsec$ \\
\xmmone & $05^{h}33^{m}44.4^{s}$& $-24\degr10\arcmin48\arcsec$\\
\hline
\cltwo& $05^{h}33^{m}39.2^{s}$& $-24\degr12\arcmin 58\arcsec$ \\
R clump &$05^{h}33^{m}38.0^{s}$& $-24\degr12\arcmin 59\arcsec$\\
\xmmtwo & $05^{h}33^{m}37.8^{s}$&$-24\degr12\arcmin50\arcsec$\\
\hline 
  \end{tabular} 
  \end{flushleft} 
\label{tab:pos}
  \end{table}

\begin{figure}[h]
\begin{center}
\epsfig{file=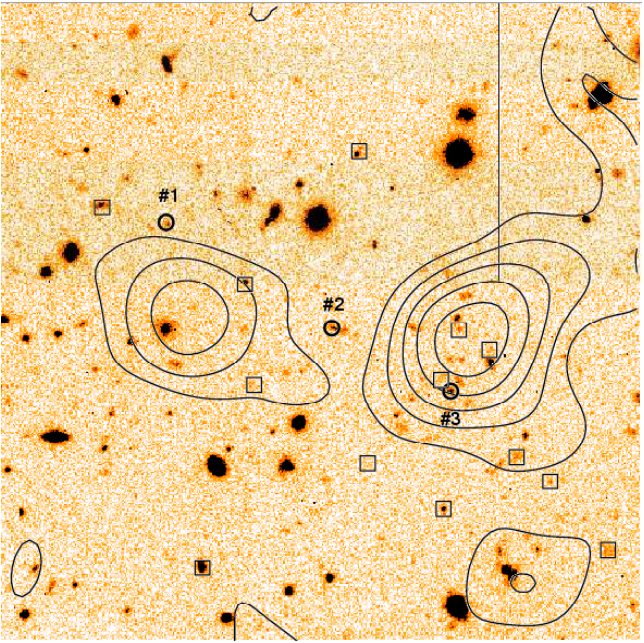,width=8.5cm} \caption{Zoom on \cltwo. 
The contours of the X-ray emission are overlaid over the VLT/FORS2 image
taken in R--band.  The exposure of the VLT-image is 10 min.  Same
symbols as in Fig.~\ref{fig:K}.}
\label{fig:Rzoom}
\end{center}
\end{figure}

Fig.\ref{fig:K} shows the X-ray contours in sigma over the K-band
image obtained with SOFI/NTT. Fig.~\ref{fig:Rzoom} shows a zoom on
\cltwo, where the X--ray contours are overlaid on the R-band
VLT/FORS2 image. The position of the galaxies with same $J-K$ ($J-K=2-2.4$) 
color are indicated by squares and the three \cltwo galaxies with
measured redshifts by circles.   

Significant X-ray emission is detected at both cluster locations.  The
peak positions are given in Table~\ref{tab:pos}, together with the
optical center of the cluster candidates estimated from a 3-D analysis
of the galaxy concentrations using the IJK imagery data.  A $6\sigma$
compact \xray source (\xmmone) is detected only $15 \arcsec$ away from
the optical center of \clone.  For \cltwo, a possibly more extended
\xray source (\xmmtwo) is detected at the $5\sigma$ level, with a
faint extension in the North-West direction.  However this extension
coincides with bright foreground galaxies and might be due to
contamination.  Just roughly 40 arcsec east of \xmmtwo there is another 
$3 \sigma$
detection (indicated with an ``S'' in Fig.\ref{fig:K}). The estimated optical center of \cltwo is located in
between these two X--ray peaks, with an offset with respect to \xmmtwo
peak of $29\arcsec$.  However, the deeper R-band VLT image reveals a
concentration of faint galaxies (hereafter called ``R clump''), centered on 
the galaxy \#3 at
$z=1.298$ (right circle in Fig.\ref{fig:K} and Tab.\ref{tab:pos}), only $10\arcsec$ from the
X--ray peak with 5 $\sigma$ detection.

\subsection{Source extent}

In order to determine the origin of the X--ray emission, we  examine
in the following the extent of \xmmone and \xmmtwo by analyzing their
surface brightness profiles.  We excise all sources detected above
$3\sigma$ in the extraction region aperture, in particular the
secondary X-ray source found east of \xmmtwo (indicated with an 
``S'' in Fig.\ref{fig:K}).

\begin{figure}[ht]
\begin{center}
\epsfig{file=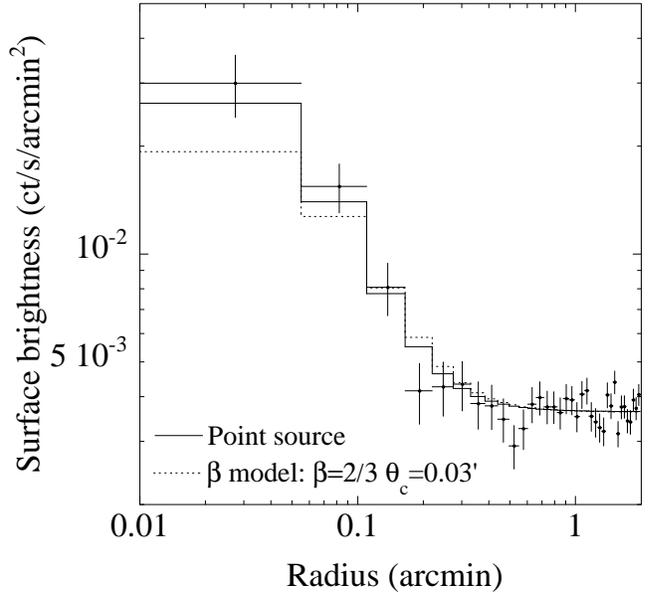,width=8.5cm} \caption{Surface
brightness profile of \xmmone.  The profile is summed over MOS1 and
MOS2.  The error bars for the surface brightness are 1 $\sigma$ uncertainties. 
The full line corresponds to the
point source model by Ghizzardi (2001) plus background.  The dotted
line is a beta-model convolved with the PSF of XMM-Newton with
$\beta=2/3$ and $\Theta_c=0.026'$.  This model is marginally
consistent with the data (see text).}
\label{fig:fig3}
\end{center}
\end{figure}

\begin{figure}[ht]
\begin{center}
\epsfig{file=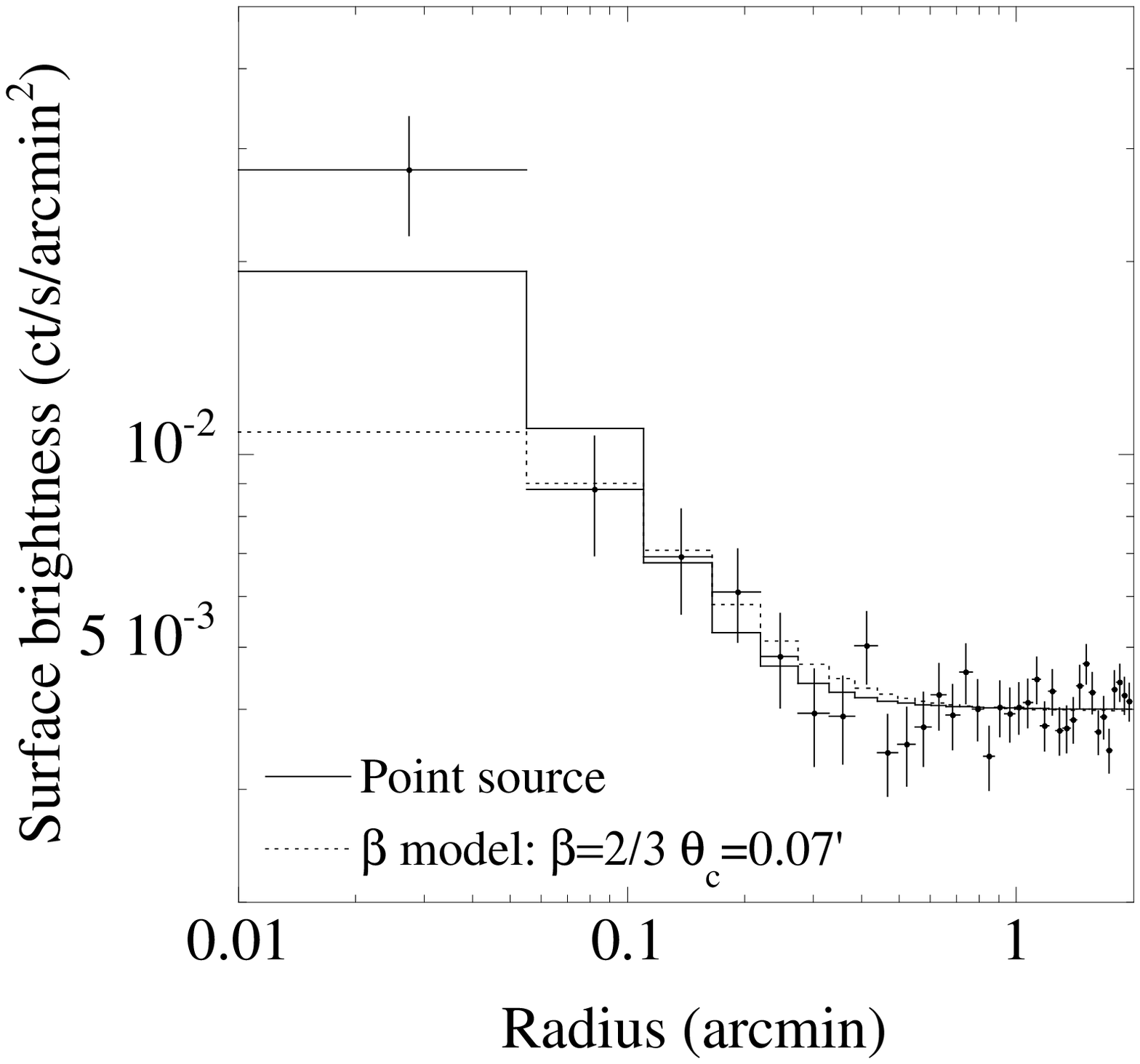,width=8.5cm} \caption{Surface
brightness profile of \xmmtwo.  The profile is summed over MOS1 and
MOS2. Error bars and full line are identical to Fig.3.
 The dotted line is a
beta-model convolved with the PSF of XMM-Newton with $\beta=2/3$ and
$\Theta_c=0.073'$.  This model is marginally consistent with the data
(see text).}
\label{fig:fig4}
\end{center}
\end{figure}

\subsubsection{\xmmone}

Fig.\ref{fig:fig3}.  shows the azimuthally averaged MOS1\&2 surface
brightness profile of \xmmone.  The center was fixed at the peak of
the X-ray emission (Table~\ref{tab:pos}). To obtain the merged surface 
brightness profile of the source we extract first the surface brightness
profile for each exposure (performed in revolution 332 and revolution 335)
and each camera (MOS1 and MOS2). 
Then we merge the surface brightness
profiles of the two exposures for each camera individually together and finally
add the profiles of the two cameras together.

In order to check whether this source is extended we compared the surface
brightness profile with the PSF model (Ghizzardi \cite{ghi01})
and adding a flat background.  The results are shown in
Fig.~\ref{fig:fig3}.  The derived $\chi^{2}$ is $36.5$ for $34$ degrees of
freedom (d.o.f.)
indicating that the emission is consistent with that of a point
source.  Fitting a \betamodel instead (with $\beta=2/3$) convolved
with the PSF does not improve the quality of the fit.  The best fit is
actually obtained for a point-like source emission ($\Theta_c=0$).  We
derived an upper limit for the core radius of $\Theta_c=0.026'$ at the
$99 \%$ confidence level ($\Delta\chi^2=6.635$).  The corresponding
profile is plotted as a dotted line in
Fig.~\ref{fig:fig3}.  $0.026'$ corresponds to roughly $14~{\rm
h_{65}^{-1}~kpc}$ at $z=1.3-1.7$.

\subsubsection{\xmmtwo}

Defining the center of \xmmtwo is more complicated since the source
seems a priori to be more extended than \xmmone and somewhat fainter
(see Fig.~\ref{fig:K}).  The determination of the center is very
important since a wrong choice could lead to an artificial measured
extent when compared to a point like source.

We first extracted the profile centered at the peak of the whole
emission.  We also consider a more conservative approach, which will
maximize the profile concentration.  For each exposure (in total two)
and each camera (MOS1 and MOS2), we further extracted four profiles
around the X-ray peak.  The centers of these profiles span a quadrat
of $2\times2$ arcsec with the X-ray peak in its center.  We choose
this spacing since a small offset between center source and measured
center will not be visible within 2 arcsec, the FWHM of the \xmm
mirrors being in the order of 4-6 arcsec.  Among the 5 resulting
surface brightness profiles per exposure and camera we choose the
profile with the highest central intensity and then merge the
different brightness profiles together.  The derived profile is
consistent with the profile centered at the overall X--ray peak but is
more noisy.  Since the two profiles are consistent, we consider in the
following only the latter profile extracted using the overall center.

The surface brightness profile together with the best fit surface
brightness profile of a point source model including background is
shown in Fig.~\ref{fig:fig4}.  Again a good fit is obtained with a
derived $\chi^{2}$ of $34.3$ for $34$ d.o.f. Fitting a \betamodel
convolved with the PSF to the brightness profile gives again a best
fit for a point source.  The 99\% confidence level for $\Theta_c$ in
this case is $\Theta_c<0.073'$ for $\beta=2/3$.  This corresponds to
$40~{\rm h_{65}^{-1}~kpc}$ at $z=1.3$.

\subsection{Count rate estimates}
\label{sec:cr}

The total MOS1\&2 count rate was estimated by integrating the surface
brightness profiles, after subtracting the best fit background level. 
The integration radius was chosen to be $20\arcsec$ to optimize the S/N
ratio.  However, this is smaller than the size of the PSF and we have
to correct for the flux loss outside the integration radius.  Since
the sources are not resolved, the correction factor was estimated from
the Encircled Energy Fraction within $20\arcsec$ for the PSF model
($EEF \sim 74\%$).  For  \xmmone (close to \clone)
we derived a PSF corrected  count rate in the
$0.3-3$~keV energy band of $(1.3\pm 0.2) \times 10^{-3}$~ct/s.  The
count rate for \xmmtwo (close to \cltwo)
is $(9.6\pm2.1) \times 10^{-4}$~ct/s.

\subsection{The origin of the X--ray emission}

The XMM observations of both \clone and \cltwo clearly reveal X-ray
emission.  However, the extracted surface brightness profiles of the
X--ray sources are comparable with the shape of the PSF and are
therefore not resolved.  Consequently, we cannot unambiguously
determine the physical origin of the sources, especially since little
is known on X-ray clusters beyond $z=1$.

It is, however, instructive to compare the upper limit set on the
source extent ($\rc < 14~{\rm h_{65}^{-1}~kpc}$ and $40~{\rm
h_{65}^{-1}~kpc}$ at the 99\% confidence limit) with the properties of
the four $z>1$ clusters known with extended X--ray emission.  The two
RDCS clusters (Rosati et al. \cite{rosati98}) 
with relatively regular morphology,
\object{RX~J0848.9+4452} ($z=1.27$) and \object{RX~J0910+5422}
($z=1.1$) have core radii at least twice as large: $100\pm30~{\rm
h_{65}^{-1}~kpc}$ (Stanford \etal~\cite{stan01}) and $171\pm53~{\rm
h_{65}^{-1}~kpc}$ (Stanford \etal~\cite{stan02}).  Interestingly, the
other RDCS cluster, \object{CIG~J0848+4453} ($z=1.26$), was originally
discovered via a search for clusters in infrared bands (Stanford
\etal~\cite{stan97}).  This cluster has an irregular morphology
(Stanford \etal~\cite{stan01}).  Nevertheless, a \betamodel fit gives
a core radius of $\sim 200~{\rm h_{65}^{-1}~kpc}$
(Holden~\cite{holden}).  A cluster with a core radius of this size at
this redshift is resolved by XMM.  Finally, \object{RX~J1053.7+5735} ($z=1.26$), discovered in deep
ROSAT observation in the Lockman hole (Hasinger \etal \cite{has98}),
was clearly resolved by XMM as a bimodal cluster (Hashimoto
\etal~\cite{hashimoto}).

A comparison can also be made with the expectation of the self-similar
model of cluster formation, where $\rc$ scales as
$\rc~\propto~\rdc~\propto~h(z)^{-1}$ (e.g. Evrard \& Gioia~\cite{eg}). 
The typical core radius of nearby clusters is $\rc \sim 0.12~\rdc$
(Neumann \& Arnaud~\cite{neu99}) where $\rdc = 2.84~(T/10~{\rm
keV})^{1/2}~{\rm h_{65}^{-1}~Mpc}$ (Evrard, Metzler \&
Navarro~\cite{emn96}).  For a relaxed part of a distant cluster with a
temperature as low as 2 keV, we would thus expect a core radius of
$73~{\rm h_{65}^{-1}~kpc}$ ($z=1.3$) and $59~{\rm h_{65}^{-1}~kpc}$
($z=1.7$).  These core radii are again excluded by our data.

In conclusion, although we cannot totally exclude that the X--ray
emission is due to exceptionally concentrated ICM, the upper limit
found on the source extent clearly favors a point source (AGN) origin. 
Those AGNs can be either foreground/background galaxies, or members of
the galaxy overdensity.  

The latter hypothesis is a priori more likely in view of the source
locations, remarkably close to the optical center of the galaxy
concentrations, \clone and \cltwo, respectively.  We further quantify
this point.  Assuming a power law AGN emission with a photon index of
$\Gamma=2$, the measured count rate of \xmmone and \xmmtwo corresponds
to an unabsorbed flux in the $[0.5-2]$~keV energy band of $S_{1} = 3.3
\times 10^{-15}~{\rm erg/s/cm^{2}}$ and $S_{2} =2.4 \times 10^{-15}~{\rm
erg/s/cm^{2}}$, respectively.  From the deep \xmm survey of the Lockman
Hole region (Hasinger \etal~\cite{has01}), the number of sources
(essentially AGN) per square degrees above those flux limits are
$N(>S_{1})\sim 300$ and $N(>S_{2})\sim 400$.  The probability that a
serendipitous AGN at least as bright as \xmmone is found by chance
within $15\arcsec$ of the center of \clone is only $1.6\%$.  For
\xmmtwo (located at $10\arcsec$ from the center of the R clump in
\cltwo), the probability is even smaller: $1.1\%$.  If we consider a
conservative distance of $30^{\prime\prime}$ 
(about 270 kpc) to take into account
possible uncertainties on the optical center determination, these
probabilities are still below $10\%$.

We looked for possible optical counterparts, as well as radio sources
in the field.  For that purpose, we used the NRAO VLA Sky Survey, NVSS
(Condon \etal~\cite{condon}).  Within $30^{\prime\prime}$ in radius of the optical
center of both concentrations, we only found one NVSS source, NVSS
J053338-241301.  This source is located only $2^{\prime\prime}$ 
away from the galaxy
member \#3 of \cltwo, an emission line galaxy, and is probably
associated with it.  However, it is too far away from \xmmtwo
($10\arcsec$) to be its counterpart, the positional uncertainty of
EPIC/MOS being less than $5\arcsec$.  It could however contaminate the
X--ray emission.  On the other hand there are at least two photometric
members (see Fig.~\ref{fig:K}) within the error box of the X-ray
source, which could be associated with it. 
 For \xmmone there is a bright optical source at a distance of 
$5^{\prime\prime}$ of the
maximum of the X-ray emission but this source is not part of the colour 
selected galaxies. However, there are several faint R 
galaxies around the X-ray maximum
but with no photometric redshift information.

\subsection{Upper limit on the ICM Luminosity}

Since, as discussed above, it is very likely that the \xray emission
we observe is at least heavily contaminated by AGN emission, the
source count rates must be regarded as upper limits for a non resolved
ICM. However, there might still be ICM emission more diffuse than the
detected X-ray sources and hidden within the background.  We estimate
an upper limit (at the $3 \sigma$ level) on this emission by fitting
the profiles with a point source plus a beta model with a core radius
of $100~{\rm h_{65}^{-1}~kpc}$ (see above) and $\beta=2/3$.  We
obtained an upper limit of $\sim 6.2\times 10^{-4}$~ct/s, lower than the
count rates of the X--ray sources. Since these count rates are lower than the
actual measured source count rates, the observed count rates 
can thus be considered as firm upper limits on the overall ICM emission.

From the measured source count rates we derive an upper limit on the ICM
luminosities of the two cluster candidates.  To convert the actual
count rate ,$(1.3\pm 0.2) \times 10^{-3}$~ct/s for \xmmone and 
$(9.6\pm2.1) \times 10^{-4}$~ct/s for \xmmtwo ,  to
luminosity 
we used the XSPEC package and the EPIC/MOS response.  We
use an absorbed MEKAL model with an assumed temperature k$T=2$~keV, a
hydrogen column density of $N_{\rm H}=2.4\times 10^{20}~{\rm
cm^{-2}}$ (Dickey \& Lockman \cite{DL}), 
and a metal abundance of 0.3 times the solar value.  For
\clone we do not have a precise redshift estimate.  The derived
bolometric X-ray luminosities are $L_{\rm bol} = (1.8\pm0.3) \times
10^{44}~{\rm h_{65}^{-2}~erg/s}$ for $z=1.7$ and $L_{\rm bol}=
(8.4\pm1.3) \times 10^{43}~{\rm h_{65}^{-2}~erg/s}$ for $z=1.3$.  For
\cltwo we derived $L_{\rm bol}= (6.2\pm1.4) \times 10^{43}~{\rm
h_{65}^{-2}~erg/s}$.

\section{Discussion}

\subsection{The nature of the cluster candidates}

The upper limit on the ICM bolometric luminosity is low enough for
\cltwo to shed some light on the nature of this galaxy system.  We
recall that spectroscopic VLT observations indicate that this is a
real physical system, and not only a chance alignment of galaxies. 
First we note that the {\it upper limit} on the luminosity of \cltwo
is similar to the luminosity of a poor cluster like Virgo in the
nearby Universe, $\sim 4\times 10^{43}~{\rm h_{65}^{-2}~erg/s}$
(Arnaud \& Evrard~\cite{ae}).  On the other hand, the matched filter
parameter, $\Lcl$, an estimate of the effective optical luminosity, is
especially high: $\Lcl=299$.  Using at face value the
correlation\footnote{$\log(L_{\rm bol,44}) = (-6.9\pm1.1)+(3.6\pm 0.8)
\log(\Lcl)$ for $\Omega_m=1$ and $H_{0}=75$~km/s/Mpc.  For the
cosmological parameters considered here, the luminosity must be
multiplied by 2.47 at $z=1.3$.  } between \xray bolometric luminosity
and $\Lcl$ derived by Donahue \etal (\cite{don01}) from their joint
optical/\rosat survey, we would expect a bolometric luminosity of at
least $3\times 10^{45}~{\rm h_{65}^{-2}~erg/s}$.  This is twice the
luminosity of Coma, $\sim 1.4\times 10^{45}~{\rm h_{65}^{-2}~erg/s}$
(Arnaud \& Evrard~\cite{ae}).  This apparent contradiction can be
naturally alleviated if we simply assume that \cltwo is a massive
proto-cluster.  In other words, \cltwo is a rich overdensity of
galaxies (as indicated by the high $\Lcl$ value), which is still
largely in a pre-virialised state (hence a low X--ray luminosity
because the gravitational potential is not deep enough to heat all the gas at the virial temperature).  In view of the galaxy distribution, we can even
speculate further.  The probable structure members (selected on their
J-K color) seem to follow a S-W/N-E filamentary structure, with the R
clump in the middle.  This R clump could be the ``seed'' (may be
partially virialized) on to which the whole structure will finally
collapse.  To further check this hypothesis we need more spectroscopic
VLT measurements to fully assess the 3D structure of the galaxy
concentration and the dynamical state of the R clump.

Since we have neither spectroscopic redshift measurement nor richness
information on \clone, the nature of this galaxy concentration remains
entirely open.  We can only tell it is not a very massive cluster in
view of the upper limit on the X--ray luminosity.

We found evidence of AGN activity in the center of both galaxy
concentrations.  Interestingly, there is also evidence of such
activity in \object{RX~J0910+5422} ($z=1.1$), a massive cluster in a
much more advanced stage of formation.  High resolution \cha deep
observation indeed revealed the presence of 3 AGNs associated with
cluster galaxies within the central $\theta < 30^{\prime\prime}$ part of the
cluster.  This aparently common AGN activity in high z (proto)
clusters might simply reflect a high AGN activity in the past: the
luminosity function of X--ray selected AGNs shows a strong density
evolution at redshift up to $\sim 1.5$ (Miyaji \etal~\cite{miyaji}). 
AGN activity could also be further boosted in regions of high galaxy
concentration as compared to the field (environmental effects).  To
assess this issue requires statistical analysis of combined Large Scale
X-ray/optical surveys (e.g. the XMM-LSS, Pierre \etal~\cite{pierre}). 
Inversely, if AGN activity for galaxies is indeed higher in overdense regions
then looking at the surroundings of AGN's might be an interesting way to
find distant galaxy clusters.

\subsection{Searching for virialized massive high redshift clusters}

Calculating the ratio of number density of massive high to low
redshift clusters is a powerful tool for the determination of
cosmological parameters (Perrenod~\cite{perr80}).  The higher this
ratio, the lower the corresponding density parameter $\Omega_m$.  Thus,
in principle it is sufficient to detect distant massive clusters and to
compare their number density to nearby ones to determine $\Omega_m$. 
However, despite the launch of \xmm and \cha the detection of extended
X-ray emission of clusters of galaxies at redshifts beyond unity is
still very difficult.  As already mentioned, only four clusters with
$z>1$ with extended X-ray emission have been found so far, and their
corresponding X-ray luminosities lie well below $L_{\rm bol}=10^{45}$
erg/s, which makes them not very massive objects.  Up to now those
clusters were either found in deep X-ray observations (Hasinger \etal
\cite{has98}; Rosati \etal~\cite{rosati}), or via a search for
clusters in infrared bands (Stanford \etal~\cite{stan97}).  These
detection techniques require relatively long exposure times and/or the
use of the largest telescopes available with small field-of-view.  This
causes the sky coverage to be very small, which makes it difficult to
assess the number density of massive distant clusters with high
precision and which is furthermore biased by the various detection
techniques applied.

Our approach for detecting massive galaxy clusters via a search in
optical/infrared bands and the lack of finding clear evidence for ICM
emission underlines the above mentioned difficulty.  Recent studies
indicate that there is clearly a decoupling of optical and X-ray
parameters for clusters at redshifts above unity (Donahue \etal
~\cite{don02}).  There are a lot of promising distant cluster
candidates detected with high signal in optical wavelengths, which at
first sight, surprisingly, do not have a corresponding luminous X-ray
counterpart.  At redshifts below unity, the correlation between
optical and X-ray luminosity seem to match better, as was indicated
recently by Donahue \etal (\cite{don01}, \cite{don02}).

What could be a possible explanation for this?  One answer could be
that the optical/infrared indicators do not match galaxies embedded in
clusters at $z>1$.  This implies a not yet accounted for evolution of
galaxy properties.  Another explanation could be that there is no or
not substantial evolution in the galaxy physics of clusters but strong
evolution of their X-ray properties and dark matter halos around
$z=1$.  This could imply, that massive clusters and their
corresponding dark matter halos, which we observe today formed or
collapsed at redshifts close to 1.  If this is true, distant clusters
detected in X-rays at redshifts close to unity should show a high
degree of significant substructure.  In fact, all clusters clearly
detected in X-rays at these redshifts show indication of substantial
substructure, which strengthens the hypothesis of important cluster
evolution at $z=1$.  In this case galaxy overdensities found in
optical/infrared wavelength bands, such as the two cluster candidates
presented here, could show the overdensities or cluster ``seeds'',
which then collapse later into clusters detectable as extended X-ray
sources.  The galaxy density enhancements can in this case be
interpreted as a tracer for regions of future cluster formation.

\section{Conclusion}

The XMM-Newton observations of two promising high redshift ($z>1$)
cluster candidates found in optical/infra-red wavelength bands, \clone
($z=1.2-1.7$) and \cltwo ($z=1.3$), reveal clear X-ray emission. 
However, the extracted surface brightness profiles of the two corresponding
X--ray
sources, \xmmone and \xmmtwo, are comparable with the shape of the
PSF. We argue that the X--ray sources are very likely AGN members belonging to
the galaxy concentrations.

We derived an upper limit for the bolometric luminosity of a hot ICM
in the range $\sim 0.7-2.1 \times 10^{44}~{\rm h_{65}^{-2}~erg/s}$ for
\clone, depending on the exact redshift.  For \cltwo the limit is
$L_{\rm bol}= (6.2\pm1.4) \times 10^{43}~{\rm h_{65}^{-2}~erg/s}$. 
This low luminosity found for a rich spectroscopically confirmed
galaxy system, can be explained if \cltwo is in fact a massive
proto-clusters which has not yet had sufficient time to collapse.

Our results indicate that rich galaxy overdensities at high redshift
found in optical/infrared surveys are not necessarily massive
virialised clusters.  
On the other hand they are of unique value for our understanding of
structure formation and the dynamics of cluster collapse.  
Such surveys can reveal pre-virialised large
galaxy overdensities or cluster ``seeds'', which will collapse later
into clusters detectable as extended X-ray sources.  Detailed
multiwavelength follow-up of large optical surveys is essential. 
Multi-object spectroscopy is required not only to confirm the
physical reality of the systems, but to fully address their 3D
structure and the dynamical state of the structure.  Combined
observations with high resolution X-ray instruments (like \cha) and
high throughput instruments (like \xmm) are needed at the same time to
i) assess the level of AGN activity which might be high in such young
systems ii) map the virialised gaseous part of the structures.

\begin{acknowledgements}

We would like to thank S. Majerowicz for providing the code to extract surface
brightness profiles from photon-event tables from \xmm data.

\end{acknowledgements}

\end{document}